\documentclass[%
reprint,
superscriptaddress,
 amsmath,amssymb,
 aps,
prb,
]{revtex4-2}

\usepackage{graphicx}
\usepackage{dcolumn}
\usepackage{bm}

\usepackage[locale=US,separate-uncertainty=true]{siunitx}
\usepackage[english]{babel}
\usepackage{mhchem}
\usepackage{float}
\usepackage{braket}
\usepackage{accents}
\usepackage{mathtools}
\usepackage{makecell}
\usepackage{colonequals}

\newcommand{\hhbar}{\text{\raisebox{.45ex}{\rotatebox{15}{--}}\hspace{-0.5em}$h$}}
\newcommand{\thefontsize}{The current font size is: \f@size pt}

\makeatletter
\newsavebox\myboxA
\newsavebox\myboxB
\newlength\mylenA
\newcommand*\xoverline[2][0.75]{%
    \sbox{\myboxA}{$\m@th#2$}%
    \setbox\myboxB\null
    \ht\myboxB=\ht\myboxA%
    \dp\myboxB=\dp\myboxA%
    \wd\myboxB=#1\wd\myboxA
    \sbox\myboxB{$\m@th\overline{\copy\myboxB}$}
    \setlength\mylenA{\the\wd\myboxA}
    \addtolength\mylenA{-\the\wd\myboxB}%
    \ifdim\wd\myboxB<\wd\myboxA%
       \rlap{\hskip 0.5\mylenA\usebox\myboxB}{\usebox\myboxA}%
    \else
        \hskip -0.5\mylenA\rlap{\usebox\myboxA}{\hskip 0.5\mylenA\usebox\myboxB}%
    \fi}
\makeatother

\newcommand{\bunderline}[1]{\underline{#1\mkern-4mu}\mkern4mu }

\begin{document}

\preprint{APS/123-QED}

\title{Effective long-range attraction of moiré excitons under the influence of atomic reconstructions and anisotropic screening}

\author{Nils-Erik Schütte}
\affiliation{Carl von Ossietzky Universität Oldenburg, Fakultät V, Institut für Physik, 26129 Oldenburg, Germany}

\author{Carl Emil Mørch Nielsen}
\affiliation{Institute of Physical Chemistry, University of Hamburg, 22607 Hamburg, Germany}

\author{Niclas Götting}
\affiliation{Carl von Ossietzky Universität Oldenburg, Fakultät V, Institut für Physik, 26129 Oldenburg, Germany}

\author{Alexander Steinhoff}
\affiliation{Carl von Ossietzky Universität Oldenburg, Fakultät V, Institut für Physik, 26129 Oldenburg, Germany}

\author{Gabriel Bester}
\affiliation{Institute of Physical Chemistry, University of Hamburg, 22607 Hamburg, Germany}

\author{Christopher Gies}
\affiliation{Carl von Ossietzky Universität Oldenburg, Fakultät V, Institut für Physik, 26129 Oldenburg, Germany}



\begin{abstract}

    The moiré pattern, which emerges due to a relative rotation between two monolayers of transition metal dichalcogenides, features a long lattice period for small twist angles.
    The resulting band structure modulation acts as an effective potential for interlayer excitons (IXs), which can realize correlated many-body phenomena.
    Here, we aim for a material-realistic modelling of the exciton-exciton interaction, taking into account lattice reconstructions and an exciton-exciton potential that incorporates the highly anisotropic screening imposed by the two-dimensional bilayer and the dielectric background.
    We find strong modifications of the on-site interaction induced by the change of the moiré potential during lattice reconstructions, while for long-range interactions on the length scale of the moiré period, anisotropic dielectric screening leads to a crossover from a repulsive to an attractive interaction.
    The interaction potential and hopping amplitudes serve as parameters for a Bose-Hubbard model on the moiré lattice, which we use to explain correlated behavior of interlayer excitons.

\end{abstract}

\maketitle


\section{Introduction \label{sec_introduction}}

Stacking two monolayers of transition metal dichalcogenides (TMDs) with a relative twist results in a moiré pattern that extends over many conventional unit cells.
This superlattice opens up the possibility for investigating many-body quantum phenomena, such as quantum phase transitions and correlated states \cite{qi2025competition, kennes2021moire, wilson2021excitons}.
Underlying these observations is often a periodic band gap modulation.
Charge carriers and excitonic complexes experience this modulation as an effective potential with the periodicity of the moiré lattice \cite{wu2017topological,wu2018theory}.
 As this situation of particles in a periodic lattice is similar to a Hubbard model, studies investigated the Fermi- \cite{hubbard1963electron} as well as the Bose-Hubbard (BH) model \cite{gersch1963quantum} for moiré systems \cite{gotting2022moirebosehubbard, tang2020simulation, gao2024excitonic, pan2020quantum, morales-duran2022nonlocal}.
In conjunction with the theoretical predicitions of the BH model, experiments on moiré TMDs find mainly a Mott-insulating phase for excitons \cite{gao2024excitonic, fowler-gerace2024transport, gu2022dipolar}, but also preliminary evidence of a superfluid phase has been shown \cite{fowler-gerace2024transport}.
Next to those, a series of other phases can be realized such as excitonic insulators \cite{ma2021strongly}, quantum Hall insulators \cite{qi2025competition} and generalized Wigner crystals \cite{huang2021correlated}.

Considering interlayer excitons (IXs), the reason for the formation of correlated insulating states is the dipole-dipole-like repulsive interaction between these composite bosons \cite{park2023dipole, li2020dipolar}. However, a material-realistic description of the exciton-exciton interaction that takes the anisotropic screening of the bilayer TMD into account, is missing.
Current models of IX interactions are mostly limited to the simplest form of a dipole-dipole potential, where dielectric screening, if at all, is only included as effective dielectric constant of the background. Thus, the subtle interplay between in-plane and out-of-plane interactions caused by the highly anisotropic screening in 2D layers is neglected.
Furthermore, in the regime of small twist angles atomic reconstructions strongly alter the ideal moiré lattice \cite{li2021lattice,nielsen2023accurate, rodriguez2023complex}.
This reordering within the moiré unit cell has direct impact on IXs via a change in the moiré potential \cite{zhao2024tunable, li2022morphology} that has not yet been explored in detail.

Here, we consider a \ce{MoS2}/\ce{WS2} heterobilayer with small relative twist angles of $\SI{1}{\degree}-\SI{3.9}{\degree}$.
By applying a low-energy continuum model for excitons that are localized in the minima of the periodic moiré potential, we calculate the Wannier functions taking into account atomic reconstructions modeled by accurate force-field relaxations of the moiré unit cell.
We map the localized excitons onto an extended BH model.
To determine the on-site Coulomb interaction $U$ and the nearest-neighbor interaction $V$ between excitons, we solve Poisson's equation including the anisotropic screening of the bilayer itself as well as a background of surrounding dielectric media.
Computing the Hubbard parameters and solving the BH model in mean-field approximation enables us to determine the quantum phases of interlayer excitons in the system.

We find that, compared to the unscreened rigid model, the on-site interaction is drastically influenced by atomic reconstructions resulting in an increased on-site interaction for large twist angles, while suppressing it for smaller twist angles.
While the lattice reordering has vanishing influence on inter-site interactions, the dielectric screening strongly modifies long-range interactions between excitons, resulting in a crossover from a repulsive to an attractive interaction at certain distances, which can further be tuned by the dielectric environment.

The paper is organized as follows.
Sec. \ref{sec:model} introduces the applied model including the representation of the moiré potential and the excitonic Hamiltonian.
Sec. \ref{sec:dft_and_force_field_relaxation_of_twisted_tmd_bilayers} briefly describes the lattice reordering in the moiré unit cell due to atomic reconstructions and introduces the used force-field relaxations.
In Sec. \ref{sec:effects_of_atomic_reconstructions} we discuss the change of excitonic properties due to atomic reconstructions by comparison with the rigid model.
In Sec. \ref{sec:moire_bose_hubbard_model} we present the BH model for IXs and the calculation of the hopping parameters, while the calculation of the interaction parameters is given in Sec. \ref{sec:exciton_exciton_interaction}.
Sec. \ref{sec:quantum_phases} presents the quantum phases of excitons predicted by the BH model.
Finally, we summarize our results in a conclusion.

\section{Model} \label{sec:model}

We consider a \ce{MoS2}/\ce{WS2} heterobilayer with the $\text{R}_{\text{X}}^{\text{X}}$-stacking structure (see Ref. \cite{tong2017topological} for a definition of the used naming convention of stacking structures).
After twisting, the stacking configuration varies across the moiré unit cell.
Considering rigidly rotated monolayers, this variation is extremely smooth.
However, the pristine moiré pattern changes due to atomic reconstructions.

TMD heterobilayers, such as \ce{MoX2}/\ce{WX2} (X $=$ S, Se) feature a type-II band alignment at the $K$-point in reciprocal space due to the weak hybridization \cite{chen2016ultrafast}.
Upon photoexcitation of intralayer electron-hole pairs,
efficient interlayer charge transfer leads to a spatial separation of electrons and holes.
The strong Coulomb interaction leads to the formation of interlayer excitons \cite{schmitt2022formation}, where the electrons are located in the \ce{MoS2} layer, while the holes remain in the \ce{WS2} layer.
Due to the twist, the $K$-points of the respective monolayers are slightly shifted.
This increases the radiative lifetime of excitons \cite{choi2021twist,huang2022excitons}, allowing them to thermalize before recombination.

The local band gap of the bilayer system depends on the local stacking configuration which varies through the moiré unit cell.
The resulting band gap modulation acts as an effective potential for IXs \cite{wu2017topological}.
The modulation of the binding energy is neglected, because it varies only marginally across the lattice \cite{wu2018theory}.
We apply a continuum model, where it is assumed that the stacking configuration and the local band gap are smooth functions in the moiré unit cell \cite{wu2018theory, wu2017topological}.\\
As input to our model, we use the $K$-$K$ band gap, obtained from small untwisted unit cell density functional theory (DFT) calculations for different layer stackings, which then are mapped onto the moiré unit cell \cite{nielsen2023accurate}.
The key approximation is that the calculated $K$-$K$ gaps are direct, while in the twisted system, the gap is indirect due to a shift between the $K$-points of the monolayers.
However, since the considered twist angles are small in our case, the shift of the $K$-points is also modest, which justifies this approximation.
To obtain a continuous variation of the band gap, it is interpolated over the unit cell.
The moiré potential can be fitted with a Fourier series expansion \cite{wu2018theory, wu2017topological}:
\begin{align} \label{eq_moire_potential_fourier_expansion}
    V^{\text{M}} (\bm{r}) = \sum_{\bm{G}^{\text{M}}} V_{\bm{G}^{\text{M}}} \text{e}^{\text{i} \bm{G}^{\text{M}} \bm{r}} .
\end{align}
Here, the sum runs over moiré reciprocal lattice vectors $G^{\text{M}}$ from several shells in the reciprocal lattice.
This is due to the fact that the moiré potential of the reconstructed system strongly differs from a sine-like curve, as will be shown in Sec. \ref{sec:dft_and_force_field_relaxation_of_twisted_tmd_bilayers}.
For an accurate fit eight shells of reciprocal lattice vectors are used.
The moiré potential is a real-valued quantity and exhibits three-fold rotational symmetry.
Because of this, all Fourier coefficients corresponding to reciprocal lattice vectors, which constitute a hexagon, can be represented by $V_{\bm{G}^{\text{M}}} = V \text{e}^{\pm \text{i} \theta}$ with alternating sign of the phase $\theta$, if the coefficients are sorted by the angle of the corresponding reciprocal lattice vectors.

We approximate the excitons as bosons by neglecting their inner fermionic structure \cite{lohof2023confinedstate}.
This is justified by the fact that the Bohr radius is small compared to the unit cell of the moiré lattice.
Therefore, we describe the center-of-mass (COM) motion of IXs in the moiré potential landscape with an effective single-particle Hamiltonian
\begin{align} \label{eq:effective_single_particle_hamiltonian}
    H_0 = - \frac{\hhbar^2}{2M} \Delta _{\bm{r}} + V^{\text{M}} (\bm{r})
\end{align}
with the total mass $M = m_{\text{e}} + m_{\text{h}}$ of the IXs.

\section{DFT and force-field relaxation of twisted TMD bilayers} \label{sec:dft_and_force_field_relaxation_of_twisted_tmd_bilayers}

At small twist angles, the pristine moiré pattern of TMD bilayers is changed by atomic reconstructions due to interatomic forces.
For TMD bilayers with $\text{R}_{\text{X}}^{\text{X}}$-- stacking, the $\text{R}_{\text{M}}^{\text{X}}$-- and $\text{R}_{\text{X}}^{\text{M}}$--configurations are favored by the system because they lower the stacking energy \cite{carr2018relaxation, enaldiev2020stacking}.
These areas grow in size during a relaxation process, while the regions with an $\text{R}_{\text{X}}^{\text{X}}$-stacking shrink.
The shrinkage occurs because the $\text{R}_{\text{X}}^{\text{X}}$-configuration maximizes the stacking energy \cite{rosenberger2020twist}.
Atomic reconstructions are accompanied by an increase of the energy due to the resulting strain, leading to a competition between the interatomic forces and the induced strain.
For small twist angles/large moiré unit cells, the layers are nearly aligned inducing a minimal strain.
For small unit cells, a very large strain is needed which inhibits the formation of large areas with the same stacking.
The described lattice reordering starts to be significant for angles below $\SI{4}{\degree}-\SI{5}{\degree}$ \cite{nielsen2023accurate, rodriguez2023complex, zhao2023excitons}.

While it is not possible to apply ab initio DFT simulations directly to the moiré unit cells, which can contain more than 100000 atoms, we employ a classical force-field for the geometry optimization.
Our used procedure is elaborated in Ref. \cite{nielsen2023accurate}, and briefly summarized in the following.
Important for TMD bilayer systems is the correct modelling of the interlayer van der Waals (vdW) interaction.
As proposed in Ref. \cite{naik2019kolmogorov}, a Kolmogorov-Crespi (KC) potential can be used to describe the vdW interaction.
Parameters of the KC potential (lattice constant, intra- and interlayer distance) are fitted to DFT relaxations for small untwisted systems.
In the special case of lattice-matched bilayers it is sufficient to fit the KC parameters only for the six high-symmetry stacking configurations $\mathrm{R}_{\mathrm{X}}^{\mathrm{X}}$/$\mathrm{R}_{\mathrm{M}}^{\mathrm{M}}$, $\mathrm{R}_{\mathrm{M}}^{\mathrm{X}}$, $\mathrm{R}_{\mathrm{X}}^{\mathrm{M}}$, $\mathrm{H}_{\mathrm{X}}^{\mathrm{M}}$/$\mathrm{H}_{\mathrm{M}}^{\mathrm{X}}$, $\mathrm{H}_{\mathrm{X}}^{\mathrm{X}}$, $\mathrm{H}_{\mathrm{M}}^{\mathrm{M}}$, which retain the $C_3$ rotational symmetry.
This is due to the fact that every point in the moiré unit cell of a lattice-matched system can be represented by the stacking configuration of the bilayer system with a specific translation between the monolayers but without a relative twist.
For large moiré periods, the local stacking configuration varies continuously through the unit cell but the stacking configuration at each point can be approximated by a superposition of the high-symmetry stacking configurations.

Since the moiré potential is obtained from DFT calculations as well, at this point, we also describe its calculation using a grid-based interpolation method.
For each transition metal atom $m$ in the bottom layer, a shift $\Delta r_{m} = (\Delta x, \Delta y)_{m}$ can be defined as the in-plane distance to the nearest transition metal atom in the top layer.
$\Delta r_{m}$ is adjusted relative to the twist of the two monolayers.
$\mathrm{R}_\mathrm{X}^\mathrm{X}$ is represented by $\Delta r_{m} = 0$.
Then, DFT calculations are performed for each value of $\Delta r_{m}$ to get the local band gap at each point of the transition metal atoms of the bottom layer, i.e.
\begin{align} \label{eq_band_gap}
    E_{\mathrm{g}} = E_{\mathrm{g}} \left( \Delta r_{m}, d (\Delta r_{m}) \right) ,
\end{align}
where $d(\Delta r_{m})$ is a shift-dependent interlayer spacing.
To obtain a continuous band gap variation, $E_{\mathrm{g}}$ is interpolated over the whole moiré unit cell.

We use the local band gap at the $K$-valley, i.e. the energy difference between the local extrema of the valence band in \ce{WS2} and the conduction band in \ce{MoS2}.
As previously explained, this band gap of untwisted DFT calculations serves as approximation for the $K$-$K$ gap of the twisted heterostructure.

\section{Effects of atomic reconstructions on moiré interlayer excitons} \label{sec:effects_of_atomic_reconstructions}

\begin{figure}
    \centering
    \includegraphics[scale=1]{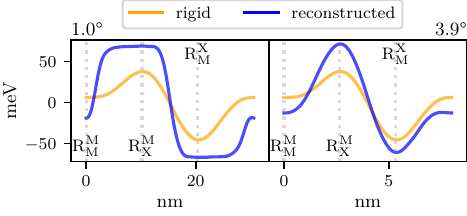}
    \caption{Fourier fit of the moiré potential along the diagonal of the moiré unit cell.
    Potential is shown for twist angles of \SI{1}{\degree} and \SI{3.9}{\degree} and for the rigid and reconstructed unit cell.}
    \label{fig:moire_potentials_diagonal}
\end{figure}

We now use the moiré potential obtained from the DFT calculation as ingredient for the effective single-particle Hamiltonian from Eq. \eqref{eq:effective_single_particle_hamiltonian}.
Before solving the eigenvalue problem, it is instructive to study the changes of the moiré potential due to atomic reconstructions.
Fig. \ref{fig:moire_potentials_diagonal} compares the moiré potential of the rigid and relaxed system for two twist angles along the diagonal through the moiré unit cell.
Due to lateral atomic reconstructions, the potential adopts the triangular shape of the lattice for small twist angles.
The potential minima and maxima are much broader than in the rigid case.
Along the diagonal, the potential has a quantum well-like shape, which has a direct impact on the excitonic dispersion.
For larger twist angles, the induced strain increases, making the lateral reconstructions less significant and we find similar behavior of the moiré potential at \SI{3.9}{\degree}.
For all angles, out-of-plane reconstructions lead to a much deeper moiré potential of $\sim \SI{135}{meV}$, compared to $\sim \SI{80}{meV}$ in the rigid system, which is in good agreement with previous estimates in the literature of 100-$\SI{260}{meV}$ \cite{guo2020shedding, lu2019modulated, yu2017moire}.

We solve the eigenvalue problem given by the Hamiltonian \eqref{eq:effective_single_particle_hamiltonian} in a Bloch basis:
\begin{align} \label{eq:bloch_functions}
    \chi_{\bm{Q}}^{(\alpha)} (\bm{r}) = \frac{1}{\sqrt{\mathcal{A}}}\sum_{\bm{G}^{\text{M}}} c_{\bm{G}^{\text{M}}}^{(\alpha)} (\bm{Q}) \text{e}^{\text{i} (\bm{Q} - \bm{G}^{\text{M}}) \bm{r}} ,
\end{align}
where $\bm{Q}$ is the in-plane quasi-momentum of the COM motion of IXs, and $\mathcal{A}$ is the crystal area.

\begin{figure}
    \centering
    \includegraphics[scale=1]{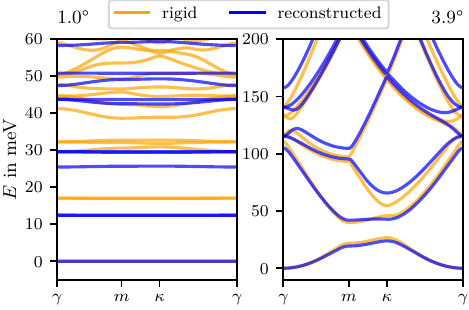}
    \caption{Excitonic band structure of the rigid and reconstructed system for twist angles of \SI{1}{\degree} and \SI{3.9}{\degree}.}
    \label{fig:band_structure}
\end{figure}

\begin{figure}
    \centering
    \includegraphics[scale=1]{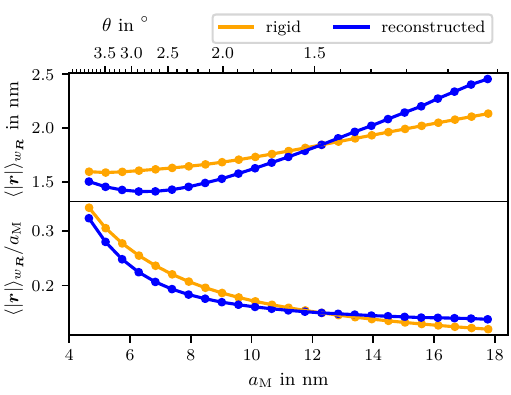}
    \caption{Weighted average extent (top) and relative extent (bottom) of the excitonic Wannier functions in dependence of the twist angle (moiré period).}
    \label{fig:wf_localization_extent}
\end{figure}

The resulting excitonic dispersion is shown in Fig. \ref{fig:band_structure}.
For small twist angles we observe a flat dispersion in both systems, which was already found in rigid systems \cite{brem2020tunable}.
The flat bands correspond to IXs with a high effective mass, i.e. inert particles leading to a reduced affinity of the excitons to move across the lattice.
In the BH model in Sec. \ref{sec:moire_bose_hubbard_model}, the mobility of the IXs is quantitatively captured by the hopping amplitude.
Since the relaxed moiré potential has deeper, broader minima, much like a wider quantum‑well trap, the resulting excitonic sub‑bands lie closer together. On the other hand, in the rigid system with deeper, narrower minima the bands are pushed further apart.
This could make excited states more accessible at non-zero temperature.
For larger twist angles the bands exhibit a more pronounced curvature in both cases, reflecting the nearly identical shape of the underlying potentials in this range.
The quantum‑well‑like shape of the moiré potential, together with its depth \cite{tran2019evidence} and the reduced mobility of excitons at small twist angles, suggests that IXs become trapped in the potential minima.
Due to the localization it is natural to change from the Bloch basis to a Wannier function representation via
\begin{align} \label{eq_wannier_functions}
    w_{\bm{R}}^{(\alpha)} (\bm{r}) = \frac{1}{\sqrt{N}} \sum_{\bm{Q}} \text{e}^{\text{i} \bm{Q} \bm{R}} \chi_{\bm{Q}}^{(\alpha)} (\bm{r}) ,
\end{align}
where the Wannier state is localized at position $\bm{R}$ and $N$ is the number of in-plane quasi-momenta $\bm{Q}$.
Eq. \eqref{eq_wannier_functions} defines orbitals for the IXs seen as effective single particles.
We concentrate on the lowest band and set $\alpha = 0$ from now on.
\begin{figure}
    \centering
    \includegraphics[scale=1]{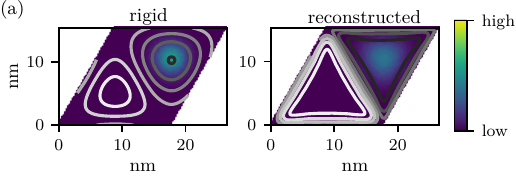}
    \includegraphics[scale=1]{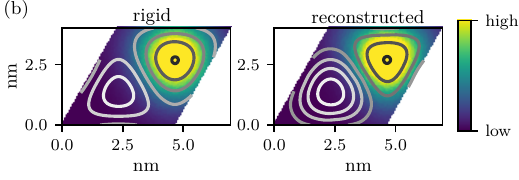}
    \caption{Ground state excitonic Wannier functions in the moiré unit cell as color map and moiré potential as contour plot for (a) \SI{1}{\degree} and (b) \SI{3.9}{\degree}.}
    \label{fig:wannier_functions}
\end{figure}
Fig. \ref{fig:wannier_functions} shows the excitonic Wannier functions for both systems at the smallest and largest twist angle.
The Wannier functions are localized in the minima of the moiré potential for all twist angles and both systems, in accordance with the above assumption and previous results \cite{jin2019observation}.
At small twist angles, the Wannier functions smear out in the broad potential minimum of the relaxed system and are less localized in comparison to the rigid case.
Furthermore, the shape of the Wannier function changes to match the triangular shape of the potential minima.
A completely opposing behavior appears at large twist angles, where the Wannier functions of the relaxed system are more strongly localized due to the deepened moiré potential.\\
Quantitative information about the localization of the Wannier functions can be obtained from the extent of the IXs.
As an absolute measure of the extent, we calculate the first absolute moment of $ \left| w_{\bm{R}} (\bm{r}) \right| ^2$:
\begin{align}
    \langle \left| \bm{r} \right| \rangle _{w_{\bm{R}}} = \int \left| \bm{r} - \bm{R} \right| \left| w_{\bm{R}} (\bm{r}) \right| ^2 \, \mathrm{d}^2 r .
\end{align}
This absolute measure of the extent is shown in Fig. \ref{fig:wf_localization_extent} as a function of the twist angle, with a smaller extent corresponding to stronger localization.
The crossover comes from the competition of the depth and the shape of the moiré potential.
Although out-of-plane lattice reconstructions induce a deeper potential at all angles, the broad shape of the minima dominates over the depth and leads to a lesser localization of the excitonic wave functions at small twist angles.
With increasing twist angle, the lateral reconstructions loose significance followed by a narrowing of the potential until the depth outweigh this effect at $\theta \approx \SI{1.47}{\degree}$.
The bottom of Fig. \ref{fig:wf_localization_extent} shows the absolute extent normalized to the moiré period $\langle \left| \bm{r} \right| \rangle _{w_{\bm{R}}} / a_{\text{M}}$, which can be viewed as relative extent.
It shows that although the moiré unit cell grows as the twist angle decreases, the Wannier functions do not expand across the cell in proportion to this growth.
We therefore expect that the on-site repulsion between excitons will generally decrease with smaller twist angles, however, not with the same rate as the area of the moiré unit cell grows.

\section{Moiré Bose-Hubbard Model} \label{sec:moire_bose_hubbard_model}

The strong spatial localization of the IX wave functions at the minima of the moiré potential allows us to map the system onto a Bose-Hubbard (BH) model \cite{gotting2022moirebosehubbard}.
We consider the extended version of the BH model by using the Hamiltonian
\begin{align}
    \begin{aligned} \label{eq:extended_bh_hamiltonian}
    \hat{H} = & - t \sum_{\langle i j \rangle} \hat{b}_{i}^{\dagger} \hat{b}_{j}^{\phantom \dagger} + \frac{U}{2} \sum_{i} \hat{n}_{i} (\hat{n}_{i} - 1) \\
    & - \mu \sum_{i} \hat{n}_{i} + \frac{V}{2} \sum_{\langle i j \rangle} \hat{n}_{i} \hat{n}_{j} ,
    \end{aligned}
\end{align}
which can be derived from a general Hamiltonian for spinless bosons on a lattice \cite{bogner2020quantum}.
Here, $\hat{b}^{\dagger}_{i}$, $\hat{b}^{\phantom \dagger}_{i}$ are the bosonic creation and annihilation operators and $\hat{n}_{i} = \hat{b}^{\dagger}_{i} \hat{b}^{\phantom \dagger}_{i}$ is the boson number operator.
$t$ denotes the nearest-neighbor hopping and $U$ and $V$ are the on-site and nearest-neighbor interactions, respectively.

The BH Hamiltonian \eqref{eq:extended_bh_hamiltonian} describes an eigenvalue problem that can be solved for a set of parameters $(t, \mu, U, V)$ independently of any material platform on which the system is implemented.
In Appendix \ref{app:bh_model} we present the solution of the extended BH model in mean-field approximation together with the resulting phase diagram.
Here, we consider an experimental setup with a fixed number of excitons in the lattice, corresponding to a canonical ensemble.
This changes the parameter space from the grand canonical ensemble $(t, \mu, U, V)$ to $(t, \rho, U, V)$, where $\rho$ is the exciton density (given as the number of excitons per moiré unit cell).
Material and platform specific properties of correlated excitonic phases enter via the model parameters $t$, $U$ and $V$.
This is where we establish a connection by using the Wannier states that were introduced in Sec. \ref{sec:effects_of_atomic_reconstructions} to determine these parameters for our specific setup.

\begin{figure}
    \centering
    \includegraphics[scale=1]{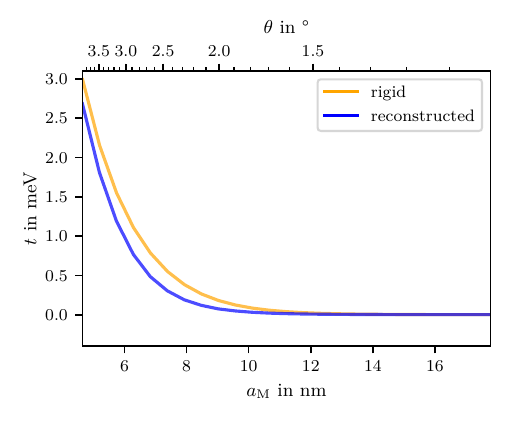}
    \caption{Nearest-neighbor hopping amplitude $t$ calculated for IXs in a \ce{MoS2}/\ce{WS2} heterobilayer in dependence on the twist angle.}
    \label{fig:hubbard_t_MoS2-WS2_R__rigid_relaxed}
\end{figure}

Generally, the hopping between $n$-th nearest-neighbor lattice sites $\bm{R}$ and $\bm{R}_{n}$ is given by:
\begin{align} \label{eq_hopping_amplitude}
    t_n &= \int w_{\bm{R}}^{*} (\bm{r}) \hat{H}_{0} (\bm{r}) w_{\bm{R}_{n}}^{\phantom *} (\bm{r}) \, \mathrm{d}^2 r \nonumber \\
    & = \frac{1}{N} \sum_{\bm{Q}} \mathrm{e}^{\mathrm{i} \bm{Q} (\bm{R} - \bm{R}_{n})} E_{\bm{Q}} .
\end{align}
We neglect $t_0$ because it just merely represents a constant energy.
Eq. \eqref{eq_hopping_amplitude} is effectively the Fourier transform of the lowest band and, therefore, all information of the hopping amplitude is already incorporated in the IX dispersion.
Results of the hopping parameters as a function of the twist angle are shown in Fig. \ref{fig:hubbard_t_MoS2-WS2_R__rigid_relaxed}.
As mentioned in the last section, at small twist angles the affinity of the IXs to move across the lattice is reduced because of the flat dispersion, which was already found in Ref. \cite{brem2020tunable}.
This results in vanishingly small hopping parameters, where no difference between the rigid and relaxed system can be seen.
As the twist angle increases, the hopping amplitudes also increase due to a more pronounced curvature of the dispersion and differences due to atomic reconstructions become apparent.
However, by further increasing the twist angle, these differences vanish again due to a lesser influence of atomic reconstructions.
Overall, the hopping amplitudes of the relaxed system are slightly smaller due to the flatter bands induced by the quantum well-like shape of the potential.
We note that the next-nearest-neighbor hopping is small which is why we neglect it.
The calculation of the Hubbard $U_n$ parameters is much more elaborate, as we show in the next section.

\section{Macroscopic modelling of the exciton-exciton interaction in two-dimensional heterobilayers} \label{sec:exciton_exciton_interaction}

\begin{figure}
    \centering
    \includegraphics[scale=1.2]{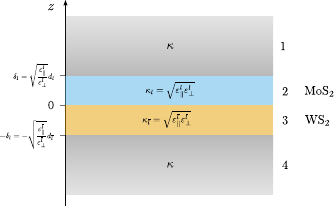}
    \caption{Schematic illustration of a TMD heterobilayer encapsulated in a dielectric environment.}
    \label{fig:dielectric_screening_schematic}
\end{figure}
A key characteristic of atomically thin crystals is the predominant role of interaction effects.
The IX-IX interaction in 2D bilayers is often modeled by a classical dipole-dipole potential, while dielectric screening of the environment or the bilayer itself is neglected or only considered via a dielectric constant, rather than a non-local function \cite{park2023dipole, li2020dipolar, kremser2020discrete}.
This completely neglects the anisotropic dielectric response, which becomes especially important for long-range interactions.
We account for the anisotropy by calculating macroscopic non-local dielectric functions for the in-plane and out-of-plane direction corresponding to intra- and interlayer Coulomb interactions.
As shown in detail in the Appendix \ref{app:poisson_solution}, we do this by solving Poisson's equation for a situation depicted in Fig. \ref{fig:dielectric_screening_schematic}.
The TMD bilayer with layers labeled by $l$ and $\bar{l}$ is encapsulated between bulk \ce{hBN}.
Dielectric constants are denoted by $\kappa, \kappa_l, \kappa_{\bar{l}}$, which are given by the geometric mean of the respective in-plane ($\parallel$) and out-of-plane ($\perp$) dielectric constants: $\kappa_l = \sqrt{\varepsilon_{\parallel}^{l} \varepsilon_{\perp}^{l}}$ and analogously for $\kappa, \kappa_{\bar{l}}$.
The thickness of the layers are denoted by $d_l$, $d_{\bar{l}}$.


The result of our calculation is the potential energy between two excitons, which consists of four contributions:
\begin{align}
    V_{\text{IX}} (q) & = V_{ll}(q) + V_{\bar{l} \bar{l}}(q) - 2 V_{l \bar{l}}(q) \nonumber \\
    & = \frac{W^{\text{intra}}(q)}{\varepsilon_{ll}(q)} + \frac{W^{\text{intra}}(q)}{\varepsilon_{\bar{l} \bar{l}}(q)} - \frac{2 W^{\text{inter}}(q)}{\varepsilon_{l \bar{l}}(q)} \label{eq:ix_potential_contributions} ,
\end{align}
where $W^{\text{intra}}(q) = \frac{e_0^2}{2 \varepsilon_0 \mathcal{A} q}$ is the bare Fourier-transformed 2D electrostatic potential and $W^{\text{inter}}(q) = \frac{e_0^2}{2 \varepsilon_0 \mathcal{A} q} \text{e}^{- \frac{q}{2} (d_l + d_{\bar{l}})}$ contains an exponential decay due to the interlayer separation of electrons and holes.
Here, $e_0$ is the elementary charge.
The first two terms are given by $V_{ll}(q)$ and $V_{\bar{l} \bar{l}}(q)$, which model the intralayer interactions between electrons in layer $l$ and holes in layer $\bar{l}$.
$V_{l \bar{l}}(q)$ represents the two possible interlayer interactions between an electron in layer $l$ and a hole in layer $\bar{l}$ ($V_{l \bar{l}}(q)$ being symmetric in $l$ and $\bar{l}$).
For the dielectric functions we find analytical expressions which are provided in Appendix \ref{app:poisson_solution}.

Since our goal is to calculate the Hubbard $U_n$ parameters as expectation values of the exciton-exciton potential, we need the latter in real space representation.
To improve convergence of the inverse Fourier transformation, we account for the finite out-of-plane extent of electron and hole wave functions in the intralayer Coulomb integrals via an exponentially decaying form factor $F_l(q)$ as described in Ref. \cite{villafane2023twistdependent}.
The inverse Fourier transformation is performed numerically to obtain $V_{\text{IX}}(r)$:
\begin{align}
    V_{\text{IX}}(r) &= \frac{1}{(2 \pi)^2} \int_{\mathbb{R}^2} (F_l V_{ll} + F_{\bar{l}} V_{\bar{l} \bar{l}} - 2 V_{l \bar{l}}) \text{e}^{\text{i} \bm{r} \bm{q}} \, \text{d}^2 \bm{q} \nonumber \\
    &= \frac{1}{(2 \pi)^2} \int_{0}^{\infty} \bigg[
    \begin{aligned}[t]
        & (F_l V_{ll} + F_{\bar{l}} V_{\bar{l} \bar{l}} - 2 V_{l \bar{l}}) q \\
        & \times \int_{0}^{2 \pi} \text{e}^{\text{i} r q \cos (\vartheta)} \, \text{d} \vartheta \bigg] \text{d} q
    \end{aligned}\\
    &= \frac{1}{2 \pi} \int_{0}^{\infty} (F_l V_{ll} + F_{\bar{l}} V_{\bar{l} \bar{l}} - 2 V_{l \bar{l}}) q J_0(q r) \, \text{d} q ,
\end{align}
where $J_0$ is the 0-th order Bessel function of the first kind.
We omitted the $q$-dependencies of the form factor and the potentials for clarity.

\begin{figure}
    \centering
    \includegraphics[scale=1]{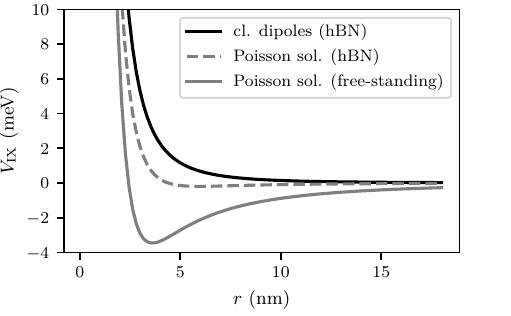}
    \caption{Exciton-exciton interaction potential modeled as classical dipole-dipole interaction (solid black) with hBN as homogeneous dielectric background, or calculated as the solution of the Poisson equation for an anisotropic heterostructure either freestanding in vacuum (solid gray) or encapsulated in \ce{hBN} (dashed gray).}
    \label{fig:exciton_potential_real_space}
\end{figure}
To demonstrate the significant changes of the IX-IX potential, we show $V_{\text{IX}}(r)$ for a heterobilayer \ce{MoS2}/\ce{WS2} in comparison with the classical dipole-dipole potential in Fig. \ref{fig:exciton_potential_real_space}.
As expected, there are moderate differences between the potential which accounts for the anisotropic topology and the potential which assumes a homogeneous dielectric background, due to the two different physical situations which they describe.
In the limit of vanishing distance between the excitons, both potentials behave similarly and diverge as $r \rightarrow 0$.
For larger interaction distances we see qualitative differences between both models.
While the dipole-dipole potential is repulsive for all $r$, anisotropic screening effects lead to a sign change of the potential.
At first glance, this might seem unintuitive for dipoles that are parallely aligned.
However, it can be interpreted with the anisotropic screening in the following way.
Since the in-plane dielectric constant of TMD monolayers is larger than the out-of-plane dielectric constant \cite{laturia2018dielectric}, in-plane interactions are screened more efficiently than out-of-plane interactions.
While the repulsive electron-electron and hole-hole interactions in layers $l$ and $\bar{l}$ are purely in-plane, the attractive interactions between electrons and holes have also an out-of-plane component.
Since this out-of-plane component is suppressed less strongly by screening, there is a certain distance $r^{*}$, at which the in-plane repulsive interactions are sufficiently screened such that the electron-hole interactions take over and lead to an effective attraction.
Coincidentally, the magnitute of $r^{*}$ lies in the range of the moiré lattice constant at small twist angles.
This opens up the possibility to interesting correlated physics in moiré structures, which can be tuned by the twist angle as well as by the dielectric environment.

\begin{figure}
    \centering
    \includegraphics[scale=1]{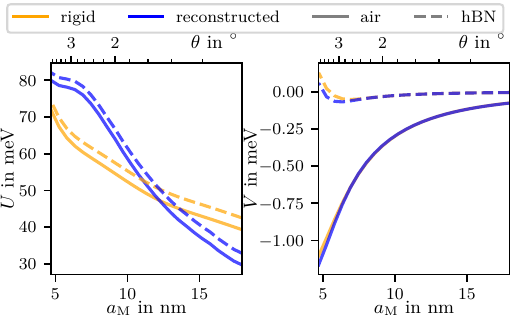}
    \caption{Hubbard $U_n$ parameters: on-site interaction $U$ (left) and nearest-neighbor interaction $V$ (right).
    Parameters are calculated for a \ce{MoS2}/\ce{WS2} heterobilayer freestanding in vacuum (solid) and encapsulated in \ce{hBN} (dashed).}
    \label{fig:hubbard_u_MoS2-WS2_R__rigid_relaxed}
\end{figure}

We use the screened potential to calculate the Hubbard $U_n$ parameters between excitons at $n$-th neighboring lattice sites (where $U_0 = U$ is the on-site term and $U_1 = V$ is the nearest-neighbor interaction).
$U_n$ can be calculated as the expectation value of the interaction potential
\begin{align} \label{eq:hubbard_u}
    U_{n} = \int_{\mathbb{R}^2} \left| w_{\bm{R}} (\bm{r}) \right| ^2 \left| w_{\bm{R}_{n}} (\bm{r}^{\prime}) \right| ^2 V_{\text{IX}} (| \bm{r} - \bm{r}^{\prime} |) \, \mathrm{d}^2 \bm{r} \mathrm{d}^2 \bm{r^{\prime}} ,
\end{align}
where $\bm{R}_n$ is the vector corresponding to the site, which is a $n$-th nearest-neighbor of lattice site $\bm{R}$.
The on-site and nearest-neighbor interaction as a function of the twist angle are depicted in Fig. \ref{fig:hubbard_u_MoS2-WS2_R__rigid_relaxed}.
The two different interaction regimes are dominated by changes of either the atomic reconstructions or the anisotropic screening.
The former predominantly modifies the on-site interaction in a highly twist-angle dependent way.
At small twist angles, lateral atomic reconstructions cause the Wannier functions to smear out, thereby reducing the interaction strength.
In contrast to that, the stronger (absolute) localization of the Wannier functions at larger twist angles increases the interaction.
The crossover between the two regimes appears at $\theta \approx \SI{1.5}{\degree}$ and is directly related to the crossover of the localization/extent of the Wannier functions observed in Fig. \ref{fig:wf_localization_extent}.
Generally, our calculated on-site interactions are comparable to findings in experiments of $\sim 10$ - $\SI{37}{meV}$ \cite{park2023dipole,fowler-gerace2024transport}.

Considering the nearest‑neighbor interaction in the right panel of Fig.~\ref{fig:hubbard_u_MoS2-WS2_R__rigid_relaxed}, it becomes evident that lattice reordering has a negligible effect on the long‑range excitonic interaction.
This is obvious from Eq.~\eqref{eq:hubbard_u}, where the overlap of the Wannier functions is governed primarily by the distance between lattice sites, rather than by the localization of each exciton.
In contrast to that, the dielectric screening does not only have quantitative, but also qualitative impact on the long-range interaction.
As explained earlier, the crossover of the exciton potential from a repulsive to an attractive regime appears on the same length scale as the moiré period.
While the nearest-neighbor exciton interaction is attractive at all twist angles when the bilayer is free-standing (vacuum), the interaction with an \ce{hBN} encapsulation possesses a repulsive regime at large twist angles and an attractive regime at smaller ones.
At this point it should be stressed that the nearest-neighbor interaction is one to two orders of magnitude weaker than the on-site one.
While this will make a direct measurement of the long-range interaction generally difficult, the fact that the excitonic potential has a global minimum is in agreement with findings of generalized Wigner crystals \cite{huang2021correlated}.

\section{Quantum phases of moiré excitons} \label{sec:quantum_phases}

\begin{figure}[H]
    \centering
    \includegraphics[scale=1]{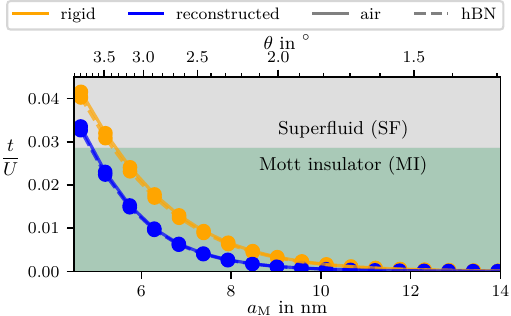}
    \caption{Quantum phases of moiré IXs in a \ce{MoS2}/\ce{WS2} heterostructure as a function of the twist angle for a unit filling factor.}
    \label{fig:phase_extended_mott}
\end{figure}

Having calculated the Bose-Hubbard parameters, we can determine the phases of the IXs as a function of the twist angle.
To reduce the influence of fermionic effects, we restrict our analysis to a fixed filling factor of one boson per lattice site ($\rho = 1$).
We find that for fillings with $\rho<1$, excitons form a superfluid phase.
The strongly suppressed long-range interactions  makes density-wave and supersolid phases inaccessible.
Phases for $\rho=1$ as a function of the twist angle are shown in  Fig.~\ref{fig:phase_extended_mott}.
Consistent with experimental results, excitons are in a Mott-insulating (MI) state for most twist angles \cite{deng2025frozen, gao2024excitonic, fowler-gerace2024transport}.
Due to the increased on-site interaction at large twist angles, atomic reconstructions push the phases deeper into the MI region.
However, also for the reconstructed system a phase transition to the superfluid phase is predicted at large twist angles.
Generally, the superfluid phase has been predicted for (moiré free) TMD heterostructures \cite{ma2021strongly, fogler2014hightemperature} and has recently been measured in a \ce{MoSe2}/\ce{WSe2} bilayer \cite{cutshall2025imaging}. For moiré superlattices, experimental results also hint towards a superfluid phase of interlayer excitons \cite{fowler-gerace2024transport}.

\section{Conclusion} \label{sec:conclusion}

We have examined properties of interlayer excitons in the moiré lattice of a twisted \ce{MoS2}/\ce{WS2} heterostructure under the influence of atomic reconstructions that become particularly relevant at small twist angles.
The lattice reordering is modeled with a classical force-field relaxation and captured by a modified moiré potential for interlayer excitons in terms of a low-energy continuum Hamiltonian. The resulting wave functions for excitons trapped in the moiré potential are mapped to an extended Bose-Hubbard lattice model.
To obtain a material-realistic exciton-exciton interaction potential accounting for the highly anisotropic screening in few-layer TMD systems, we explicitly solve Poisson's equation for a macroscopic model of the dielectric heterostructure.


We find that the interaction potential between interlayer excitons in TMD bilayers exhibits a non-intuitive crossover from a repulsive interaction for short distances to an attractive long-range interaction, which is unexpected from aligned dipolar excitons in free space.
This results in a repulsive on-site interaction, which is drastically influenced by the lattice reordering.
At large twist angles, the repulsion increases while it decreases at small ones.
Atomic reconstructions have little effect on the nearest‑neighbor interaction, but we find the anisotropic dielectric environment to play a substantial role.
Since the transition from a repulsive to an attractive interaction appears on the same length scale as the moiré lattice period, the nearest-neighbor can be  attractive or repulsive via its dependence on the dielectric environment.

Finally, we used the numerically obtained Hubbard parameters to determine the phase of the excitons as a function of the twist angle.
At fractional lattice fillings, the excitons form a superfluid due to the weak nearest-neighbor interaction, making the density-wave phase or the supersolid phase inaccessible.
At unit filling factor, excitons form a Mott-insulating state at most twist angles in agreement with findings in the literature.

Our work improves the theoretical understanding of moiré interlayer excitons by including accurate lattice reconstructions modifying the moiré potential and using a realistic description of the exciton-exciton Coulomb interaction.
Our observation that the exciton-exciton potential undergoes a crossover from repulsive short-range to attractive long-range interactions is of particular interest for experiments in 2D excitonic lattices.
For example, a recent experiment \cite{deng2025frozen} reported frozen exciton dynamics in a \ce{WS2}/\ce{WSe2} heterostructure for a duration of $\sim \SI{80}{ns}$, explained by a stabilization of the Mott-insulating phase due to long-range repulsive exciton interactions. The crossover to an attractive interaction potential could cause a similar effect, which may be worth investigating for different TMD material combinations.

An attractive exciton-exciton interaction has been deduced from measured exciton line shifts in a free-standing \ce{WSe2}/\ce{WS2} bilayer \cite{sun2022enhanced}.
While in this work, it was attributed to the exchange interaction between excitons, our finding of an attractive interaction due to anisotropic screening could be an alternative explanation for this phenomenon, especially for the free-standing heterobilayer setup of Ref. \cite{sun2022enhanced} for which our model predicts an enhanced attraction compared to a setup with dielectric encapsulation.


\vspace{1cm}

\paragraph*{Acknowledgements}
The authors would like to thank Frederik Lohof for many useful discussions.
The project has been funded by Deutsche Forschungsgemeinschaft (DFG) within the Priority Program SPP2244 via the projects Gi1121/4-2 and STE 2943/1-2

\clearpage
\onecolumngrid

\appendix


\section{Extended Bose-Hubbard model} \label{app:bh_model}

The fact that the Hamiltonian (\ref{eq:extended_bh_hamiltonian}) includes products of annihilation and creation operators at neighboring lattice sites (bilocal operators) makes it impossible to solve the corresponding eigenvalue problem for an infinite lattice exactly \cite{gheeraert2016MeanField}.
We use the mean-field (MF) approximation to find a numerical solution to the extended BH Hamiltonian.
A derivation can be found in Ref. \cite{may2013hardcore}, which we will sketch in the following.

If we write the bosonic operators as their mean value plus a small deviation term:
\begin{align}
    & \hat{b}_{i} = \varphi_{i} + \delta \hat{b}_{i} \\
    & \hat{b}_{i}^{\dagger} = \varphi_{i}^{*} + \delta \hat{b}_{i}^{\dagger} \\
    & \hat{n}_{i} = \varrho_{i} + \delta \hat{n}_{i}
\end{align}
the MF theory approximates the product of operators at different lattice sites by a sum of products containing only the mean values and local operators:
\begin{align}
    \hat{b}_{i}^{\dagger} \hat{b}_{j}^{\phantom \dagger} & \approx - \varphi_{i}^{*} \varphi_{j}^{\phantom *} + \varphi_{i}^{*} \hat{b}_{j}^{\phantom \dagger} + \varphi_{j}^{\phantom *} \hat{b}_{i}^{\dagger} \\
    \hat{n}_{i} \hat{n}_{j} & \approx - \varrho_{i} \varrho_{j} + \varrho_{i} \hat{n}_{j} + \varrho_{j} \hat{n}_{i} .
\end{align}
With these approximations, the MF Hamiltonian reads:
\begin{align}
    \begin{aligned} \label{eq:mf_hamiltonian_lattice}
    \hat{H} \approx \hat{H}^{\text{MF}} = & - t \sum_{i} (- \varphi_{i}^{*} \xoverline{\varphi}_{i}^{\phantom *} + \xoverline{\varphi}_{i}^{*} \hat{b}_{i}^{\phantom \dagger} + \xoverline{\varphi}_{i}^{\phantom *} \hat{b}_{i}^{\dagger}) + \frac{U}{2} \sum_{i} \hat{n}_{i} (\hat{n}_{i} - 1) - \mu \sum_{i} \hat{n}_{i} \\
    & + \frac{V}{2} (- \varrho_{i} \xoverline{\varrho}_{i} + 2 \xoverline{\varrho}_{i} \hat{n}_{i}) .
    \end{aligned}
\end{align}

\begin{figure}[h]
    \centering
    \includegraphics{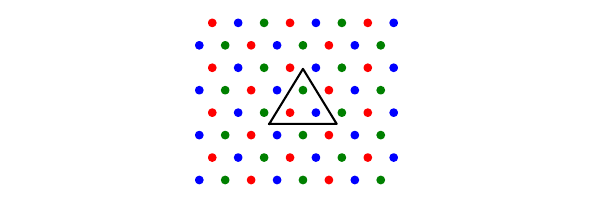}
    \caption{Triangular lattice with three sublattices, marked by different colors.
    The black lines mark the unit cell of the lattice.}
    \label{fig:triangular_sublattice}
\end{figure}

Solving the standard BH model in MF approximation would require only a diagonalization for one lattice site.
However, a Hamiltonian with a nearest-neighbor interaction requires to distinguish between different lattice sites.
In order to do so, the whole lattice is split into different sublattices.
Instead of diagonalizing Eq.~(\ref{eq:mf_hamiltonian_lattice}) for a single lattice site, it will be solved for a unit cell.
The unit cell is defined to be the set of single lattice sites corresponding to different sublattices, and which can be repeated by translation to reconstruct the whole lattice.
In Fig. \ref{fig:triangular_sublattice} we show the sublattices and the unit cell for the triangular lattice.
Now, the Hamiltonian for a single unit cell $\hat{H}_{\mathrm{UC}}^{\mathrm{MF}}$ can be written as a sum over the sublattices $X$:
\begin{align}
    \begin{aligned} \label{eq:mf_hamiltonian_sublattice}
    \hat{H}_{\mathrm{UC}}^{\mathrm{MF}} = & - t \sum_{X} (- \varphi_{X}^{*} \xoverline{\varphi}_{X}^{\phantom *} + \xoverline{\varphi}_{X}^{*} \hat{b}_{X}^{\phantom \dagger} + \xoverline{\varphi}_{X}^{\phantom *} \hat{b}_{X}^{\dagger}) + \frac{U}{2} \sum_{X} \hat{n}_{X} (\hat{n}_{X} - 1) - \mu \sum_{X} \hat{n}_{X} \\
    & + V \sum_{X} \xoverline{\varrho}_{X} (\hat{n}_{X} - \frac{1}{2} \varrho_{X}) .
    \end{aligned}
\end{align}
In principle, $\xoverline{\varphi}_{X}$ and $\xoverline{\varrho}_{X}$ are still sums over all nearest-neighbors of the lattice site corresponding to sublattice $X$.
However, here they will be expressed as summations over sublattices, instead of sums over individual lattice sites.
Therefore, the so-called \textit{adjacency matrix} $\bunderline{\bm{N}}_{X Y}$ is introduced.
It expresses how many lattice sites of sublattice $Y$ are the neighbors of the lattice site belonging to sublattice $X$.
Here, three sublattices will be used, for which $\bunderline{\bm{N}}$ is:
\begin{align}
    \bunderline{\bm{N}} = \begin{pmatrix}
        0 & 3 & 3 \\
        3 & 0 & 3 \\
        3 & 3 & 0
    \end{pmatrix} .
\end{align}
With this adjacency matrix, $\xoverline{\varphi}_{X}$ and $\xoverline{\varrho}_{X}$ can be expressed as
\begin{align}
    & \xoverline{\varphi}_{X} = \sum_{Y} \bunderline{\bm{N}}_{X Y} \varphi_{Y} \\
    & \xoverline{\varrho}_{X} = \sum_{Y} \bunderline{\bm{N}}_{X Y} \varrho_{Y} .
\end{align}
For reference, the quantities $\varphi$ and $\varrho$ without any index refer to the mean values:
\begin{align}
    & \varphi = \frac{1}{n_{\mathrm{Lattice}}} \sum_{X} \varphi_{X} \\
    & \varrho = \frac{1}{n_{\mathrm{Lattice}}} \sum_{X} \varrho_{X} ,
\end{align}
where $n_{\mathrm{Lattice}}$ is the number of sublattices.

\begin{figure}[h]
    \centering
    \includegraphics[scale=0.8]{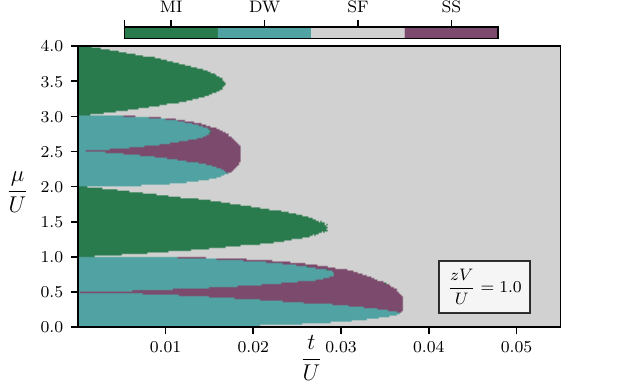}
    \caption{Phase diagram of the extended Bose-Hubbard model solved in mean-field approximation.}
    \label{fig:phase_diagram_extended}
\end{figure}

\section{Parameters and material constants} \label{app:parameters_and_material_constants}

In Table \ref{tab:parameters_and_material_constants} we list the parameters and material constants which we used for our simulations.
\begin{table}[h]
    \centering
    \begin{tabular}{l|l|c|c}
        quantity    &   calculation & value &   reference \\ \hline
        $\kappa$ of \ce{hBN}    &   $\kappa = \sqrt{\varepsilon_{\parallel}^{\ce{hBN}} \varepsilon_{\perp}^{\ce{hBN}}}$ &   3.76    &   \cite{segura2018natural} \\
        $\kappa_l$ of \ce{MoS2} &   $\kappa = \sqrt{\varepsilon_{\parallel}^{l} \varepsilon_{\perp}^{l}}$   &   9.69    &   \cite{laturia2018dielectric} \\
        $\kappa_{\bar{l}}$ of \ce{WS2} &   $\kappa = \sqrt{\varepsilon_{\parallel}^{\bar{l}} \varepsilon_{\perp}^{\bar{l}}}$   &   10.64    &   \cite{laturia2018dielectric} \\
        $d_l$ of \ce{MoS2}  &   &   \SI{0.618}{nm}  &   \cite{kylanpaa2015binding} \\
        $d_{\bar{l}}$ of \ce{WS2}  &   &   \SI{0.6219}{nm}  &   \cite{kylanpaa2015binding} \\
        $m_{\text{e}}$  &   &   $0.42 m_0$  &   \cite{wu2018theory} \\
        $m_{\text{h}}$  &   &   $0.34 m_0$  &   \cite{wu2018theory}
    \end{tabular}
    \caption{List of parameters and material constants which we used for the numerical simulations.
    $m_0$ is the mass of a free electron.}
    \label{tab:parameters_and_material_constants}
\end{table}

\section{Solution of Poisson's equation with an anisotropic dielectric constant} \label{app:poisson_solution}

Poisson's equation accounting for an anisotropic dielectric constant is given by:
\begin{align}
    \nabla_{\bm{r}} \cdot \int \bar{\bar{\varepsilon}} (\bm{r}, \bm{r}^\prime) \nabla_{\bm{r^\prime}} \phi (\bm{r}^\prime) \, \text{d}^3 r^\prime = - \frac{\varrho (\bm{r})}{\varepsilon_0} ,
\end{align}
where $\bar{\bar{\varepsilon}} (\bm{r}, \bm{r}^\prime)$ is the non-local dielectric tensor.
Our goal is to find the screened electrostatic potential $\phi(\bm{r})$ between charges in a vertical dielectric heterostructure as depicted in Fig. \ref{fig:dielectric_screening_schematic}. To this end, we solve Poisson's equation for each layer of the heterostructure, respectively, and connect the individual solutions by boundary conditions that follow from Maxwell's equations.
Due to the geometry of the problem, cylindrical coordinates $\bm{r}=(x,y,z)=(\rho \textrm{cos}\varphi, \rho \textrm{sin}\varphi,z) $ are best suited to derive the screened potential.
We assume that in each layer the dielectric function is local with respect to the out-of-plane coordinate $z$ and non-local with respect to in-plane coordinates,
$\bar{\bar{\varepsilon}} (\bm{r}, \bm{r}^\prime) = \bar{\bar{\varepsilon}} (\bm{\rho}, \bm{\rho}^\prime) \delta(z-z^\prime)$, where the two-dimensional vector $\bm{\rho}$ compasses the coordinates $\rho$ and $\varphi$. Moreover, we assume rotational invariance, which removes the dependence of all quantities on the angle $\varphi$.
With this, Poisson's equation becomes:
\begin{align} \label{eq:poisson_epsilon_in_plane}
    \nabla_{\bm{r}} \cdot \int \int \bar{\bar{\varepsilon}} (\bm{\rho}, \bm{\rho}^\prime) \delta (z-z^\prime) \left( \frac{\partial \phi (\rho^\prime, z^\prime)}{\partial \rho^\prime} \bm{e}_{\rho^\prime} + \frac{\partial \phi (\rho^\prime, z^\prime)}{\partial z^\prime} \bm{e}_{z^\prime} \right) \, \text{d}^2\bm{\rho^\prime} \text{d}z^\prime = - \frac{\varrho (\rho, z)}{\varepsilon_0} .
\end{align}
Here, $\bm{e}_{\bm{\rho}^\prime}, \bm{e}_{z^\prime}$ are basis vectors whose components are given by $e_{i, \alpha} = \braket{\alpha | i (\bm{r})}, \, \alpha = \{x, y, z\}$.
Latin characters $i,j$ indicate the components of basis vectors in cylindrical coordinates.
The dielectric tensor may be represented in cylindrical coordinates as:
\begin{align}
    \bar{\bar{\varepsilon}} (\bm{r}, \bm{r}^\prime) = \sum_{i,j} \ket{i(\bm{r})} \underbrace{\bra{i(\bm{r})} \bar{\bar{\varepsilon}} (\bm{r}, \bm{r}^\prime) \ket{j(\bm{r}^\prime)}}_{\varepsilon_{ij} (\bm{r}, \bm{r}^\prime)} \bra{j(\bm{r}^\prime)} ,
    \label{eq:tensor_cyl}
\end{align}
where the position dependence of the basis vectors is taken into account.
We assume that the dielectric tensor is diagonal in cartesian coordinates:
\begin{align}
    \bar{\bar{\varepsilon}} (\bm{r}, \bm{r}^\prime) = \begin{pmatrix}
        \varepsilon_{\parallel} & 0 & 0 \\
        0 & \varepsilon_{\parallel} & 0 \\
        0 & 0 & \varepsilon_{\perp}
    \end{pmatrix} ,
\end{align}
where all the components are still non-local functions of the in-plane coordinates: $\varepsilon_{\parallel} = \varepsilon_{\parallel}(\bm{\rho}, \bm{\rho}^\prime)$, $\varepsilon_{\perp} = \varepsilon_{\perp} (\bm{\rho}, \bm{\rho}^\prime)$.
To transform the dielectric tensor from cartesian to cylindrical coordinates, we insert completeness relations into Eq.~(\ref{eq:tensor_cyl}). Thus, we obtain for example for the $\rho$-$\rho$-component:
\begin{alignat}{3}
    & \braket{\rho(r) | \bar{\bar{\varepsilon}} | \rho (r^\prime)} && = &&\bra{\rho (r)} \sum_\alpha \ket{\alpha} \bra{\alpha} \bar{\bar{\varepsilon}} \sum_\beta \ket{\beta} \bra{\beta} \ket{\rho (r^\prime)} \nonumber \\
    &&& = &&\braket{\rho(r) | x} \varepsilon_{\parallel} \braket{x | \rho(r^\prime)} \nonumber \\
    &&&&&+ \braket{\rho(r) | y} \varepsilon_{\parallel} \braket{y | \rho(r^\prime)} \nonumber \\
    &&&&&+ \braket{\rho(r) | z} \varepsilon_{\perp} \braket{z | \rho(r^\prime)} \nonumber \\
    &&& = &&\varepsilon_{\parallel} (\cos \varphi \cos \varphi^\prime + \sin \varphi \sin \varphi^\prime) \nonumber \\
    &&& = &&\varepsilon_{\parallel} \cos (\varphi - \varphi^\prime) \nonumber .
\end{alignat}
Plugging the cyclindrical representation of $\bar{\bar{\varepsilon}}$ into Eq. \eqref{eq:poisson_epsilon_in_plane}, using the orthonormality of basis vectors and writing the divergence explicitly in cylindrical coordinates yields:
\begin{align}
    &\nabla_{\bm{r}} \cdot \int \left(\varepsilon_{\parallel} (\bm{\rho}, \bm{\rho}^\prime) \frac{\partial \phi (\rho^\prime, z)}{\partial \rho^\prime}\Big( \cos (\varphi - \varphi^\prime)  \bm{e}_{\rho}-\sin(\varphi - \varphi^\prime)\bm{e}_{\varphi}\Big) + \varepsilon_{\perp}(\bm{\rho}, \bm{\rho}^\prime) \frac{\partial \phi (\rho^\prime, z)}{\partial z^\prime} \bm{e}_{z}\right) \, \text{d}^2 \bm{\rho^\prime} \nonumber \\
    & = \int_0^\infty \int_0^{2\pi} \bigg[
    \begin{aligned}[t]
    & \frac{1}{\rho} \frac{\partial}{\partial \rho} \left( \rho \cos (\varphi - \varphi^\prime) \varepsilon_{\parallel} (\bm{\rho}, \bm{\rho}^\prime) \frac{\partial \phi (\rho^\prime, z)}{\partial \rho^\prime} \right) \\
    & -\frac{1}{\rho}\frac{\partial}{\partial \varphi}\left(
    \sin(\varphi - \varphi^\prime)\varepsilon_{\parallel} (\bm{\rho}, \bm{\rho}^\prime) \frac{\partial \phi (\rho^\prime, z)}{\partial \rho^\prime}
    \right)
    + \varepsilon_{\perp} (\bm{\rho}, \bm{\rho}^\prime) \frac{\partial^2 \phi (\rho^\prime, z)}{\partial z^2} \bigg] \text{d} \varphi^\prime \rho^\prime \text{d}\rho^\prime
    \end{aligned} \\
    & = -\frac{\varrho (\rho, z)}{\varepsilon_0} .
\end{align}
As further simplification, we assume the dielectric function to be local with respect to the in-plane coordinates:
\begin{align}
    \frac{1}{\rho} \frac{\partial}{\partial \rho} \left( \rho \varepsilon_{\parallel} \frac{\partial \phi (\rho, z)}{\partial \rho} \right) + \varepsilon_{\perp} \frac{\partial^2}{\partial z^2} \phi(\rho, z) = - \frac{\varrho (\rho, z)}{\varepsilon_0} \nonumber\,.
\end{align}
This allows us to Fourier transform the electrostatic potential with respect to the in-plane coordinates:
\begin{align}
   \phi(\rho, z) = \int \text{e}^{\text{i} \bm{q}\cdot \bm{\rho}} \phi(\bm{q}, z) \, \frac{\text{d}^2 \bm{q}}{(2\pi)^2} = \int \text{e}^{\text{i} q \rho \cos \varphi_q} \phi (q, z) \, \frac{\text{d}^2 \bm{q}}{(2\pi)^2} ,
\end{align}
where $\varphi_q$ is the angle between $\bm{\rho}$ and $\bm{q}$ and the rotational symmetry of the Fourier transformed potential has been used.
With this, Poisson's equation becomes \footnote{
    During the derivation, we will use the following integrals:
    \begin{align*}
        &\int_{0}^{2\pi} \text{e}^{\text{i} q \rho \cos \varphi_q} \, \text{d} \varphi_q = 2 \pi J_0 (q \rho) \\
        &\int_{0}^{2 \pi} \cos \varphi_q \text{e}^{\text{i}q \rho \cos \varphi_q} \, \text{d} \varphi_q = 2 \pi \text{i} J_1 (q \rho) \\
        &\int_{0}^{2 \pi} \cos^2 \varphi_q \text{e}^{\text{i}q \rho \cos \varphi_q} \, \text{d} \varphi_q =  \int_{0}^{2 \pi} (1 - \sin^2 \varphi_q) \text{e}^{\text{i}q \rho \cos \varphi_q} \, \text{d} \varphi_q = 2 \pi J_0 (q \rho) - \frac{2 \pi J_1(q \rho)}{q \rho} \\
        &\int_{0}^{\infty} J_n (q \rho) J_n (q^\prime \rho) \rho \, \text{d}\rho = \frac{1}{q} \delta (q - q^\prime) ,
    \end{align*}
    where $J_n$ is the $n$-th order Bessel function of the 0-th kind.
}:
\begin{align}
    &&& \int \left[ \varepsilon_{\parallel} \frac{1}{\rho} \left( \text{i} q \cos \varphi_q \text{e}^{\text{i} q \rho \cos \varphi_q} + \rho (\text{i} q \cos \varphi_q)^2 \text{e}^{\text{i} q \rho \cos \varphi_q} \right) + \varepsilon_{\perp} \frac{\partial^2}{\partial z^2} \text{e}^{\text{i} q \rho \cos \varphi_q} \right] \phi (q, z) \, \frac{\text{d}^2 \bm{q}}{(2\pi)^2} \nonumber \\
    &&&= - \int \frac{\rho(q^\prime, z)}{\varepsilon_0} \text{e}^{\text{i} q^\prime \rho \cos \varphi_{q^\prime}} \frac{\text{d}^2 \bm{q}}{(2\pi)^2} \nonumber \\
    & \Rightarrow &&\frac{1}{(2\pi)^2} \int_{0}^{\infty} \bigg[ \varepsilon_{\parallel} \underbrace{\left( \frac{\text{i} q}{\rho} 2 \pi \text{i} J_1(q \rho) - q^2 \left( 2 \pi J_0(q\rho) - \frac{2 \pi J_1(q \rho)}{q \rho} \right) \right)}_{\substack{= 2 \pi \left( - \frac{q}{\rho} J_1(q \rho) - q^2 J_0(q \rho) + \frac{q}{\rho} J_1(q \rho) \right) \\ = - q^2 2 \pi J_0 (q \rho)}} + \varepsilon_{\perp} \frac{\partial^2}{\partial z^2} 2 \pi J_0(q \rho) \bigg] \phi(q, z) q \, \text{d}q \nonumber \\
    &&&= - \frac{1}{(2 \pi)^2} \int_{0}^{\infty} 2 \pi J_0 (q^\prime \rho) \frac{\varrho (q^\prime, z)}{\varepsilon_0} q^\prime \, \text{d}q^\prime \nonumber \\
    & \Rightarrow && \int_{0}^{\infty} J_0(q, \rho) \left[ -q^2 \varepsilon_{\parallel} + \frac{\partial^2}{\partial z^2} \varepsilon_{\perp} \right] \phi (q, z) q \, \text{d}q
    = - \int_{0}^\infty J_0 (q^\prime \rho) \frac{\varrho (q^\prime, z)}{\varepsilon_0} q^\prime \, \text{d}q^\prime \nonumber \\
    & \Rightarrow && \left( \varepsilon_{\perp} \frac{\partial^2}{\partial z^2} - \varepsilon_{\parallel} q^2 \right) \phi (q, z) = - \frac{\varrho (q, z)}{\varepsilon_0} \quad \nonumber \\
    & \Rightarrow && \varepsilon_{\parallel} \left( \frac{\varepsilon_{\perp}}{\varepsilon_{\parallel}} \frac{\partial^2}{\partial z^2} - q^2 \right) \phi (q, z) = - \frac{\varrho(q)}{\varepsilon_0} \delta (z-z_0) \label{eq:poisson_fourier}
\end{align}
In the last step, we assumed that the charge density is localized at $z = z_0$.
By introducing a rescaled out-of-plane coordinate
\begin{align}
    \widetilde{z}=z\sqrt{\frac{\varepsilon_{\parallel}}{\varepsilon_{\perp}}}= z\gamma
\end{align}
one can bring Poisson's equation for an anisotropic dielectric tensor to the form of an effectively isotropic system with dielectric constant $\kappa=\sqrt{\varepsilon_{\parallel}\varepsilon_{\perp}}$
\cite{mele2001screening, erkensten2021excitonexciton}:
\begin{align}
     \left(\frac{\partial^2}{\partial \widetilde{z}^2} - q^2 \right) \widetilde{\phi} (q, \widetilde{z}) = - \frac{\varrho(q)}{\varepsilon_0\varepsilon_{\parallel}} \delta ((\widetilde{z}-\widetilde{z}_0)/\gamma)=- \frac{\varrho(q)}{\varepsilon_0\varepsilon_{\parallel}}\gamma \delta (\widetilde{z}-\widetilde{z}_0)=- \frac{\varrho(q)}{\varepsilon_0\sqrt{\varepsilon_{\parallel}\varepsilon_{\perp}}}\delta (\widetilde{z}-\widetilde{z}_0)\,.
\label{eq:poisson_final}
\end{align}
This rescaling procedure can be straightforwardly extended to a heterostructure consisting of several layers. After obtaining the effective screened potential $\widetilde{\phi} (q, \widetilde{z})$ in terms of the rescaled coordinate, the solution for the anisotropic geometry at the physical coordinate is identified as $\phi(z=\widetilde{z}/\gamma)=\widetilde{\phi} (q, \widetilde{z})$ . For simplicity, we drop the tilde notation in the following as all coordinates are understood to be rescaled ones.
\\
\\We now seek the solution to Eq. \eqref{eq:poisson_final} for our specific situation of a heterobilayer embedded into a dielectric medium as depicted in Fig.~\ref{fig:dielectric_screening_schematic}, where Poisson's equation holds in each layer separately.
We assume that the charge density is located in the middle of one of the TMD layers, such that
\begin{align}
    z_0 = \frac{\delta_l}{2} \, \lor \, z_0 = - \frac{\delta_{\bar{l}}}{2} .
\end{align}
Here, $\delta_l = \sqrt{\frac{\varepsilon_{\parallel}^{l}}{\varepsilon_{\perp}^{l}}} d_l$ is the rescaled thickness of layer $l$ in an effective heterostructure.
By calculating the potential $\phi$ at $z = \frac{\delta_l}{2}$ and $z = - \frac{\delta_{\bar{l}}}{2}$ caused by the charge density at $z_0$ we can derive intra- and interlayer interaction potentials.

Considering the potential in one layer produced by the charge in the other layer, this corresponds to the homogeneous part of Eq. \eqref{eq:poisson_final} with a solution of the form:
\begin{align}
    \phi_i^{\text{hom}} (q, z) = B_i^{+} \text{e}^{qz} + B_i^{-} \text{e}^{-qz} .
\end{align}
Generally, the indices $i$ and $j$ will refer to any material layer in Fig. \ref{fig:dielectric_screening_schematic} including the dielectric environment, while the indices $l$ and $\bar{l}$ specifically denote the two TMD layers.
The inhomogeneous solution for the potential in a layer with charge density is given by:
\begin{align}
    &\phi_i^{\text{inhom}} (q, z) = B_i^{+} \text{e}^{qz} + B_i^{-} \text{e}^{-qz} + A_i \text{e}^{-q |z- z_0|} , \\
    &A_i = \frac{\varrho(q)}{2 \varepsilon_0 \kappa_i q} .
\end{align}
We determine the coefficients $B_i^+, B_i^-$ by means of the boundary conditions at the layer interfaces.
Since the potentials must vanish at $z \rightarrow \infty$, we have $B_1^+ = 0$ and $B_4^- = 0$.
At the interfaces between layers, we have the following boundary conditions (corresponding to the continuity of the tangential component of the electric field $\bm{E}$ and the normal component of the displacement field $\bm{D}$, respectively):
\begin{align}
    &\phi_i = \phi_{i + 1}, \\
    &\kappa_i \frac{\partial \phi_i}{\partial z} = \kappa_{i + 1} \frac{\partial \phi_{i + 1}}{\partial z} \quad , i = 1, 2, 3.
\end{align}
First, we consider the case where the charge is in the top layer, $z_0 = \frac{\delta_l}{2}$.
At the interface of materials 1 and 2 at $z = \delta_l$ we obtain from the boundary conditions:
\begin{align}
    &B_1^- \text{e}^{- q \delta_l} = A_2 \text{e}^{-q \frac{\delta_l}{2}} + B_2^+ \text{e}^{q \delta_l} + B_2^- \text{e}^{-q \delta_l} , \\
    &- \kappa q B_1^- \text{e}^{- q \delta_l} = \kappa_l q \left( - A_2 \text{e}^{-q \frac{\delta_l}{2}} + B_2^+ \text{e}^{q \delta_l} - B_2^- \text{e}^{-q \delta_l} \right) \quad .
\end{align}
At the interface of materials 2 and 3 at $z = 0$ we have:
\begin{align}
    &A_2 \text{e}^{- q \frac{\delta_l}{2}} + B_2^+ + B_2^- = B_3^+ + B_3^- , \\
    &\kappa_l \left( A_2 \text{e}^{-q \frac{\delta_l}{2}} + B_2^+ - B_2^- \right) = \kappa_l (B_3^+ - B_3^-) .
\end{align}
Finally, for the interface between 3 and 4 at $z = - \delta_l$ we have:
\begin{align}
    &B_3^+ \text{e}^{-q \delta_{\bar{l}}} + B_3^- \text{e}^{q \delta_{\bar{l}}} = B_4^+ \text{e}^{- q \delta_{\bar{l}}} , \\
    &\kappa_{\bar{l}} (B_3^+ \text{e}^{-q \delta_{\bar{l}}} - B_3^- \text{e}^{q \delta_{\bar{l}}}) = \kappa B_4^+ \text{e}^{- q \delta_{\bar{l}}} .
\end{align}
We summarize the six conditions into the following system of linear equations:
\begin{align} \label{eq:linear_system_of_equations}
    \underbrace{\begin{pmatrix}
        a & - \frac{1}{a} & - a & 0 & 0 & 0 \\
        - \kappa a & - \frac{\kappa_l}{a} & \kappa_l a & 0 & 0 & 0 \\
        0 & 1 & 1 & - 1 & -1 & 0 \\
        0 & \kappa_l & - \kappa_l & - \kappa_{\bar{l}} & \kappa_{\bar{l}} & 0 \\
        0 & 0 & 0 & b & \frac{1}{b} & - b \\
        0 & 0 & 0 & \kappa_{\bar{l}} b & - \frac{\kappa_{\bar{l}}}{b} & - \kappa b
    \end{pmatrix}}_{\underline{M}}
    \underbrace{\begin{pmatrix}
        B_1^- \\
        B_2^+ \\
        B_2^- \\
        B_3^+ \\
        B_3^- \\
        B_4^+
    \end{pmatrix}}_{\bm{x}}
    = A_2 \underbrace{\begin{pmatrix}
        c \\
        -\kappa_l c \\
        - c \\
        - \kappa_l c \\
        0 \\
        0
    \end{pmatrix}}_{\bm{v}} ,
\end{align}
where we used the following abbreviations:
\begin{align}
    a \coloneqq \text{e}^{-q \delta_l}, \quad b \coloneqq \text{e}^{-q \delta_{\bar{l}}}, \quad c \coloneqq \text{e}^{-q \frac{\delta_l}{2}} = \sqrt{a} .
\end{align}
Eq. \eqref{eq:linear_system_of_equations} is simplified to
\begin{align}
    \bm{x} = A_2 \underline{M}^{-1} \bm{v} = A_2 \tilde{\bm{x}} ,
\end{align}
which we solve using Mathematica.
We find for the potential in the top layer at $z=\frac{\delta_l}{2}$, which corresponds to the intralayer potential:
\begin{align}
    \begin{aligned}
    \phi^{ll}(q) &= A_2 (1 + \frac{\tilde{x}_2}{c} + \tilde{x}_3 c) \\
    & \equalscolon \frac{A_2 \kappa_l}{\varepsilon^{ll}(q)} .
    \end{aligned}
\end{align}
Similarly, for the potential in the bottom layer $z = - \frac{\delta_{\bar{l}}}{2}$, we have:
\begin{align}
    \begin{aligned}
    \phi^{l \bar{l}}(q) &= A_2 (\tilde{x}_4 \sqrt{b} + \frac{\tilde{x}_5}{\sqrt{b}}) \\
    & \equalscolon \frac{A_2 \kappa_l}{\varepsilon^{l \bar{l}}(q)} \text{e}^{- \frac{q}{2} (d_l + d_{\bar{l}})} ,
    \end{aligned}
\end{align}
which is the interlayer potential.
The electrostatic potentials produced by a charge in the bottom layer ($z_0= - \frac{\delta_{\bar{l}}}{2}$) are simply obtained by interchanging the roles of $l$ and $\bar{l}$.
The $q$-dependent dielectric functions are given by:
\begin{align}
    \epsilon_{q}^{ll} = & \frac{\text{e}^{- \delta_{\bar{l}} q} \kappa_{l}}{4 \left[ \kappa_{l} \cosh (\frac{\delta_{l} q}{2}) + \kappa \sinh (\frac{\delta_{l} q}{2}) \right]} \nonumber \\
    & \times \frac{ \kappa_{l} \left[ \text{e}^{2 \delta_{\bar{l}} q} (\kappa + \kappa_{\bar{l}})^2 - (\kappa - \kappa_{\bar{l}})^2 \right] \cosh (\delta_{l} q) + 2 \text{e}^{\delta_{\bar{l}} q} \sinh (\delta_{l} q) \left[ (\kappa + \kappa_{l})^2 \kappa_{\bar{l}} \cosh (\delta_{\bar{l}} q) + \kappa (\kappa_{l}^{2} + \kappa_{\bar{l}}^{2}) \sinh (\delta_{\bar{l}} q) \right] }
    { \kappa_{l} \cosh (\frac{\delta_{l} q}{2}) \left[ \kappa_{\bar{l}} \cosh (\delta_{\bar{l}} q) + \kappa \sinh (\delta_{\bar{l}} q) \right] + \kappa_{\bar{l}} \sinh (\frac{\delta_{l} q}{2}) \left[ \kappa \cosh (\delta_{\bar{l}} q) + \kappa_{\bar{l}} \sinh (\delta_{\bar{l}} q) \right] }
\end{align}
\begin{align}
    \epsilon_{q}^{l\bar{l}} = &\frac{ \kappa_{l} \cosh (\delta_{l} q) \left[ 2 \kappa \kappa_{\bar{l}} \cosh (\delta_{\bar{l}} q) + (\kappa^2 + \kappa_{\bar{l}}^{2}) \sinh (\delta_{\bar{l}} q) \right] + \sinh (\delta_{l} q) \left[ (\kappa^2 + \kappa_{l}^{2}) \kappa_{\bar{l}} \cosh (\delta_{\bar{l}} q) + \kappa (\kappa_{l}^{2} + \kappa_{\bar{l}}^2) \sinh (\delta_{\bar{l}} q) \right] }
    {\left[ \kappa_{l} - \kappa + \text{e}^{\delta_{l} q} (\kappa + \kappa_{l}) \right] \left[ \kappa_{\bar{l}} - \kappa + \text{e}^{\delta_{\bar{l}} q} (\kappa + \kappa_{\bar{l}}) \right]} \nonumber \\
    &\times 2 \text{e}^{\frac{q}{2} (\delta_l + \delta_{\bar{l}})} \text{e}^{-\frac{q}{2} (d_l + d_{\bar{l}})}
\end{align}
Our aim is to calculate the potential energy between two parallely aligned excitons as depicted in Fig. \ref{fig:dielectric_screening_schematic}.
The potential energy consists of four contributions (electron-electron, electron-hole, hole-electron, hole-hole), which are given by:
\begin{table}[H]
    \centering
    \begin{tabular}{c|c|l}
        particles   &   interaction &   potential energy \\ \Xhline{2\arrayrulewidth}
        ee  &   \makecell{repulsive \\ intralayer}  &   $V_{\text{ee}}(q) = - e_0 \phi_{\text{e}}^{ll}(q) \equalscolon \frac{W^{\text{intra}}(q)}{\varepsilon^{ll^{\vphantom \dagger}}(q)} \equalscolon V_{ll}(q)$ \\ \hline
        eh  &   \makecell{attractive \\ interlayer} &   $V_{\text{eh}}(q) = e_0 \phi_{\text{e}}^{l \bar{l}}(q) \equalscolon - \frac{W^{\text{inter}}(q)}{\varepsilon^{l^{\vphantom \dagger} \bar{l}}(q)} \equalscolon - V_{l \bar{l}}(q)$ \\ \hline
        he  &   \makecell{attractive \\ interlayer} &   $V_{\text{he}}(q) = - e_0 \phi_{\text{h}}^{\bar{l} l}(q) \equalscolon - \frac{W^{\text{inter}}(q)}{\varepsilon^{\bar{l} l^{\vphantom \dagger}}(q)} = - V_{l \bar{l}}(q)$ \\ \hline
        hh  &   \makecell{repulsive \\ intralayer}  &   $V_{\text{hh}}(q) = e_0 \phi_{\text{h}}^{\bar{l} \bar{l}}(q) \equalscolon \frac{W^{\text{intra}}(q)}{\varepsilon^{\bar{l} \bar{l}^{\vphantom \dagger}}(q)} \equalscolon V_{\bar{l} \bar{l}}(q)$
    \end{tabular}
\end{table}
Here, $e_0$ is the elementary charge.
The index at the dielectric potentials (either e or h) refers to the charge density $\varrho(q)$ which enters $A_2$:
\begin{align}
    \varrho(q) = \begin{cases}
        - \frac{e_0}{\mathcal{A}} \, & \text{ for electrons} \\
        \frac{e_0}{\mathcal{A}} \, & \text{ for holes}\,,
    \end{cases}
\end{align}
where $\mathcal{A}$ is the crystal area.
The total potential energy between two excitons finally reads:
\begin{align} \label{eq:exciton_potential_momentum_space_app}
    V_{\text{IX}} (q) & = V_{ll}(q) + V_{\bar{l} \bar{l}}(q) - 2 V_{l \bar{l}}(q) \nonumber \\
    & = \frac{W^{\text{intra}}(q)}{\varepsilon_{ll}(q)} + \frac{W^{\text{intra}}(q)}{\varepsilon_{\bar{l} \bar{l}}(q)} - \frac{2 W^{\text{inter}}(q)}{\varepsilon_{l \bar{l}}(q)} ,
\end{align}
as given in the main text.
The parameters $\kappa_l = \sqrt{\varepsilon_{\parallel}^{l} \varepsilon_{\perp}^{l}}$ (for the layers) and $\kappa$ (for the environment) are calculated as the geometric mean of the in-plane and out-of-plane dielectric constants.
As a sanity check, the potential becomes the classical dipole-dipole potential in the $q \rightarrow 0$ limit \cite{erkensten2021excitonexciton}:
\begin{align}
V_{\text{IX}} (q) \xrightarrow[]{q \rightarrow 0} V_{\text{dipole}} = \frac{e_0^2}{2 \varepsilon_0} \left( \frac{d_l}{\varepsilon_{\perp}^{l}} + \frac{d_{\bar{l}}}{\varepsilon_{\perp}^{\bar{l}}} \right) .
\end{align}
As shown in the main text, we Fourier transform Eq. \eqref{eq:exciton_potential_momentum_space_app} back to real space to obtain the potential energy of two interlayer excitons separated by in-plane distance $r$.
This potenital energy is shown in Fig. \ref{fig:exciton_potential_real_space} for a \ce{MoS2}/\ce{WS2} heterobilayer, either encapsulated in \ce{hBN} or freestanding (vacuum), together with the potential energy of classical dipoles in a homogeneous dielectric background taken to be \ce{hBN}.

\clearpage
\newpage
\bibliography{./reconstruction_paper.bib}

\end{document}